# Self-assembled Rhodium Nanoantennas for Single-Protein UV SERS

*Yanqiu Zou[1], Nicco Corduri[2], Francesco D'Amico[3], Karol Kolataj[2*], Huaizhou Jin[4], Zhenrong Zheng[1*], Yifan Yu[5], Jie Liu[5], Shukun Weng[6], Ali Douaki[6,7], Jerome Wenger[8], Shangzhong Jin[9*], Guillermo Acuna[2*], Denis Garoli[6,7,9*]*

[1] State Key Laboratory of Modern Optical Instrumentation, College of Optical Science and Engineering, Zhejiang University, Hangzhou 310027, China

[2] Department of Physics, University of Fribourg, Chemin du Musée 3, Fribourg CH-1700, Switzerland

[3] Elettra Sincrotrone Trieste S.C.p.A., S.S. 14 km 163,5 in Area Science Park, 34149,

Basovizza, TS, Italy

[4] Key Laboratory of Quantum Precision Measurement, College of Physics, Zhejiang University of Technology, Hangzhou, China

[5] Department of Chemistry, Duke University, Durham, NC 27708, USA

[6] Istituto Italiano di Tecnologia, via Morego 30, I-16163, Genova, Italy

[7] Dipartimento di Scienze e Metodi dell'Ingegneria, Università degli Studi di Modena e Reggio Emilia, Via Amendola 2, 42122, Reggio Emilia (Italy)

[8] Aix Marseille Univ, CNRS, Centrale Marseille, Institut Fresnel, AMUTech, 13013 Marseille, France

[9] College of Optical and Electronic Technology, China Jiliang University, Hangzhou 310018, China

Corresponding authors: Prof. Garoli Denis – denis.garoli@unimore.it; Prof. Guillermo Acuna – Guillermo.acuna@unifr.ch; Prof. Shangzhong Jin - jinsz@cjlu.edu.cn; Dr. Karol Kolataj - karol.kolataj@unifr.ch;

## ABSTRACT

Surface-enhanced Raman scattering (SERS) provides critical insights into analyte structure, dynamic processes, and intermolecular interactions at the single-molecule level. By exploiting the hotspot formation in the vicinity of plasmonic structures, SERS constitutes an established tool for fundamental biological research, particularly for early-stage disease diagnostics. In this context, the DNA Origami technique, with its high addressability, enables both the assembly of plasmonic nanostructures with nanometric accuracy, and the deterministic placement of a single analyte molecule precisely at the generated hotspot within them. To date, most DNA Origami based nanoantennas rely on gold or silver nanoparticles (NPs), whose plasmonic resonances are confined to the visible spectrum, severely limiting their use in other spectral ranges. To extend the operating range, we have recently established a robust strategy for self-assembling programmable ultraviolet (UV)-plasmonic dimer antennas using rhodium nanocubes. Herein, we leverage this tailored architecture to systematically investigate its performance for single-molecule UV-SERS. We demonstrated how biofabricated Rh-dimers can be used to detect the characteristic SERS signal of a single streptavidin molecule linked at the dimer's gap. Our results are validated through polarization dependent measurements that yield the expected signal modulation depending on the the dimer orientation only for the DNA origami with a protein at the hotspot. This work establishes a highly sensitive and polarization-tunable UV-SERS platform, laying a solid foundation for label-free optical investigation and bio-spectroscopy of individual biomolecules in the UV spectral range.

## Introduction

Surface-enhanced Raman scattering (SERS) provides non-destructive, label-free molecular fingerprinting with high-sensitivity, offering a deep understanding of biomolecular structure and dynamics[1–7]. This technique relies critically on nanoscale plasmonic hotspots that generate enhanced localized electromagnetic fields, a mechanism that has enabled the use of Raman spectroscopy for single-molecule detection[8–10]. So far, diverse SERS platforms have been extensively reported[11], with examples based on colloidal nanoparticles (NPs)[12], nanoporous systems[13], and nanostructures or nanoantennas prepared via different lithographic methods[14]. In general, conventional top-down fabrication methods often lacks nanometric precision and reproducibility to assemble hotspots with consistent and enhanced optical properties even for homogeneous substrates[13,15,16]. Furthermore, with these techniques it is extremely challenging to deterministically position an individual analyte within the hotspot.

To overcome these limitations, biofabrication based on DNA origami[3,17,18] has emerged as a powerful bottom-up assembly strategy by assembling functionalized nanoparticles with synthetic oligonucleotides into nanometric plasmonic structures[19–22]. In addition, the same addressable DNA scaffold can host a single target molecule directly within the nanogap[23], fulfilling both requirements for reliable single-molecule SERS. For instance, Kosti Tapio et al. reported a method for utilizing DNA origami to precisely position single analyte molecules into the plasmonic hotspot, thereby enabling single-molecule SERS measurements of various dyes and proteins[24]. In a similar approach, Schuknecht et al. reported DNA origami based assembly of tip-to-tip gold nanorod dimers with controlled gaps of approximately 8 nm, enabling label-free single-protein SERS by specifically capturing target proteins within the plasmonic hotspot[25]. So far, the practical implementation of such DNA-assembled nanoantennas has been limited to the visible/near-infrared (NIR) spectral range, as gold or silver NPs are predominantly employed with limited extensions into high-index electric materials[20,22,25–27]. However, this spectral range is strongly limiting, first because in the visible range the contribution from the fluorescence is typically significant, then because most biomolecules have small intrinsic Raman cross-sections in the visible and NIR regions[28]. Therefore, the use of higher excitation energy (UV radiation) can contribute to reduce the background fluorescence and can lead to UV resonance Raman enhancement thanks to the absorbance of the biomolecules in this spectral range (due to the presence of multiple electronic resonances)[29,30]. Nanostructures, and NPs in particular, based on metals such as aluminum (Al), magnesium (Mg), rhodium (Rh) and gallium (Ga) can exhibit localized surface plasmon resonances (LSPRs) in the UV or deep-UV wavelength[28,31–35]. Recent works have reported on the integration of two or more plasmonic materials in the same system introducing additional tunability in the spectroscopic properties[36,37]. For instance, in a recent research we investigated UV-SERS and plasmon-driven photochemistry using a composite platform of nanoporous Al decorated with Rh NPs[38]. While this approach provided valuable insights from averaged UV-SERS spectra, it offered limited control over the morphology, size, and spatial distribution of the Rh NPs, which constrained a detailed understanding of structure–property relationships at the nanoscale. More recently, we developed a strategy for constructing well-defined, programmable Rh nanocube (NCs) dimers on DNA origami scaffolds which can precisely host a biomolecule at the hotspot.[39] Rh NCs are an ideal UV-plasmonic structures due to their strong LSPR in the UV region and excellent aqueous stability, properties that have been directly leveraged for effective UV-SERS[40,41]. Building upon this tailored UV-plasmonic platform, here we demonstrate how this platform can be used to achieve single-protein UV-SERS detection. Furthermore, we show a pronounced synergistic enhancement effect between the aluminum film substrate and the Rh-NCs-dimer nanoantenna by using well-designed micropattern, showing that the UV-SERS spectrum of functionalized streptavidin inside the hotspot can be detected consistently only on top of aluminum film. Our observations are further confirmed by polarization resolved measurements, confirming previous simulations,[42–46] that verify the SERS signal of streptavidin is maximized when the polarization of the incident laser on sample is aligned parallel to the main axis of the NCs dimer.

# Results

## Concept and design

Figure 1a depicts the strategy to fabricate Rh NC dimers with an average interparticle distance of 10.7 ± 3.2 nm using the DNA origami technique. In brief, a rectangular DNA origami with dimensions of 180 nm × 5 nm × 15 nm (length × height × width) and a 8.5 nm wide 'mast' at the center is modified to incorporate through DNA hybridization two Rh NCs. At the dimer hotspot, which coincides with the upper part of the mast, a biotin modification enables the incorporation of a single streptavidin protein. This approach enables the fabrication of UV antennas based on Rh NC dimers with and without a single protein at the hotspot for control measurements. Further details on the DNA origami and the incorporation of Rh NCs and the single streptavidin can be found in previous works[39]. Figures 1b includes a TEM image of the dimer structures, while additional TEM micrographs are reported in Supporting Information (SI – note#1). Two sets of Rh NCs have been considered, with size of 19 and 24 nm, respectively. The UV/Vis extinction measurements revealed a broad resonance centered between 300 and 320 nm in buffer conditions, respectively for 19nm and 24 nm Rh NCs (see SI – note#2).

SERS measurements were performed using two excitation wavelengths. First, to probe the response of Rh NC dimers near their absorbance maximum, a 325 nm laser was coupled to a HORIBA Raman system capable of acquiring spectroscopic 2D maps and equipped with an ultraviolet polarizer in the optical path, as schematically shown in Fig. S3 (SI). Second, for comparison with the most commonly used excitation wavelength in UV-SERS, the nanostructures were investigated using a different setup (not capable of 2D mapping) coupled to a 266 nm laser and equipped with a half-wave plate to rotate the incident polarization (see Fig. S4, SI). The use of this second excitation wavelength allows probing single molecules at energies closer to the absorption bands of several aromatic amino acids, which exhibit characteristic absorption features at approximately 255 nm, 270 nm, and 286 nm, thereby enhancing the contribution of molecular UV absorbance.[47]

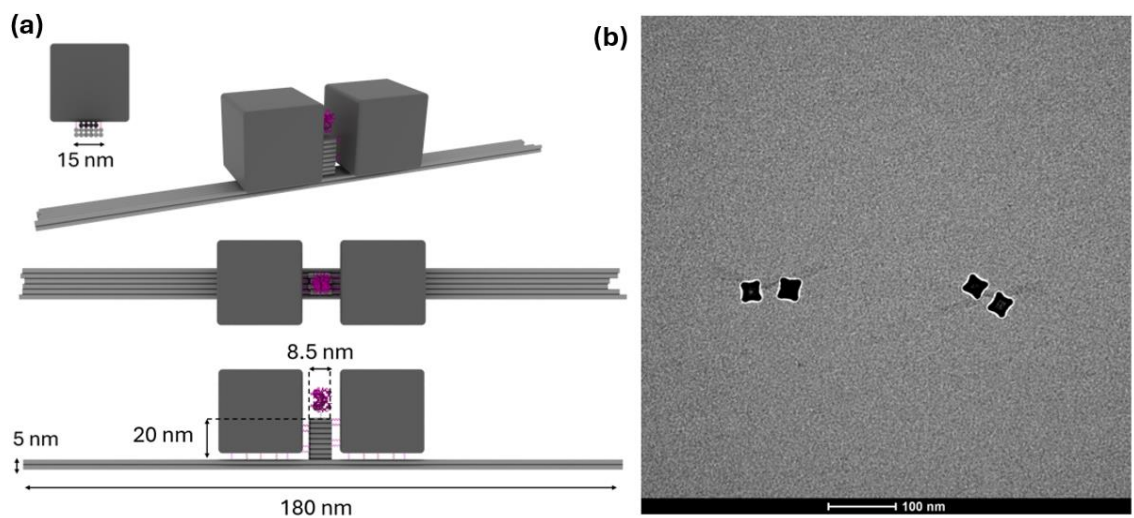

**Fig. 1** DNA Origami design with Rh NCs dimer. a) Rhodium nanocubes are precisely functionalised with DNA Origami. The mixed solution was drop-cast on top of porous aluminum films. b) TEM image of 24 nm Rh NCs dimers assembled on the DNA Origami on top of aluminum film.

**Validating single-molecule and specific DNA-Protein Binding via UV-SERS**

DNA origami was designed as a template with a length of 180 nm, 5 nm height, and a width of 15 nm with an 8.5 nm wide 'mast' placed in its center, we selected streptavidin (molecular weight ≈ 60 kDa, diameter ≈ 5 nm) as a model protein[25,48,49], and at the top of the mast, we added a biotin extension to precisely connect a single streptavidin molecule in the hotspot. This streptavidin-biotin linkage is exceptionally stable and specific, owing to a deeply buried binding pocket and cooperative conformational changes that yield ultra-high affinity[50]. To validate the specificity and single-molecule detection capability of our platform, we performed UV-SERS mapping on the DNA-origami-based Rh NC dimers in the two configurations, without and with the presence of the single streptavidin protein at the hotspot, name "streptavidin-free" and "streptavidin-functionalized" respectively. To note, while the Rh NCs dimers are prepared in solution, their deposition on the substrates used for the SERS analyses cannot be completely controlled. Therefore, the dimers can be positioned on the surface with the DNA origami laying on the surface (normal) or the dimer laying on the NPs (inverted) (see Fig. S5 – SI). Hence, it is reasonable to expect to detect SERS signals from the DNA origami from the samples prepared without the protein, while an overlap of DNA and protein vibration peaks are expected in the case of protein detection. Fig. 2 reports the UV-SERS results. We first collected comparative Raman mapping over a 70 × 70 μm² area with a step size of 14 μm (see Fig. 2(a) and (b)). The differences between the spectra collected on the samples with and without the protein are clearly observable. For these measurements we used a 1800 gr/mm grating with the minimum laser power to reduce the photodegradation. We reported on the UV-induced photodegradation in our recent works[35,38], in particular, exploring the effect of successive UV exposure on DNA and protein molecules, we were able to tune the exposure time and laser power to the values used here. The calculated average spectra are reported in Fig. 2(d) and (e). The mean spectrum from the sample without the protein at the hot-spot clearly shows several peaks associated with the DNA molecules[51]. Interesting to be noted, scanning over the sample the intensity of the Raman spectra resulted stable within a 15% variation. This is most probably due to the probed Rh NCs dimers density, where we expect at least one nanostructure per laser spot for all the tested points in the 2d-map. Fig. 2e reports the mean spectrum obtained from the streptavidin-functionalized sample. It exhibits pronounced additional stronger broad Raman bands marked with red stars. The presence of these strong, broad protein bands, alongside the persistent and sharp DNA peaks at 1061 cm$^{-1}$ (backbone) and 1548 cm$^{-1}$ (guanine), necessitates a detailed vibrational assignment to deconvolute the composite spectrum and confirm the origin of each feature. In this second case, the intensity of the Raman peaks varied significantly over the different points of the 2d-map (Fig. 2b and relative error bars in Fig. 2e). This can be reasonably justified with the high sensitivity required to detect the single small protein in the nanogap between the Rh NCs. Moreover, for statistical reasons, we expect that not all the Rh NCs dimers linked the single protein in the gap.

To proceed with the peak identification and discussion, it must be emphasized that in our previous study, we tested the UV-SERS of BSA and L-tyrosine (at high concentrations) using the same experimental system under 325 nm laser excitation[38]. The results showed that the spectra of proteins such as BSA and L-tyrosine were characterized by broad, poorly resolved features spanning approximately 1500–1700 cm$^{-1}$, with only a few discernible peaks (e.g., around 1606 cm$^{-1}$). This spectral broadening is characteristic when the laser energy is not in strong resonance with the analytes' electronic transitions, leading to a low inherent Raman cross-section and a signal dominated by a superposition of many weakly enhanced, unresolved vibrational modes, further convoluted by potential substrate-molecule interactions[52]. In contrast, the streptavidin-bound spectra presented herein, while also exhibiting broadened bands, show a set of distinct and reproducible features that enable a confident spectral deconvolution.

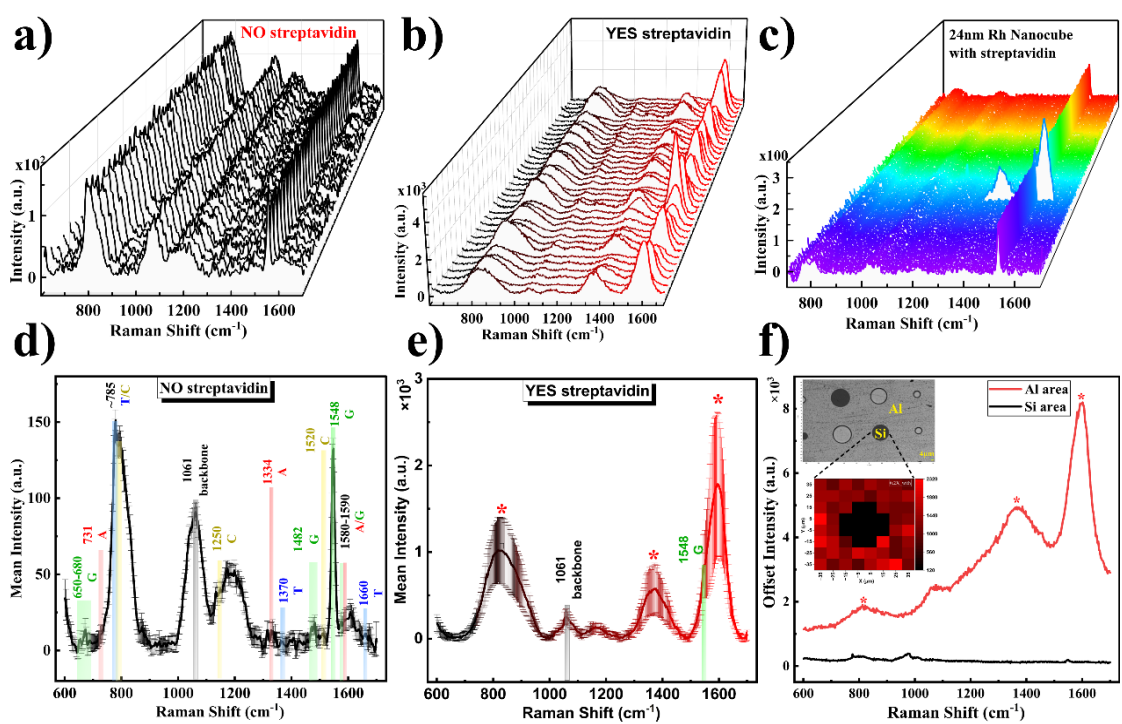

**Fig. 2.** Validation of specific DNA-protein binding and single-molecule detection via UV-SERS. UV-SERS mapping spectra obtained on aluminum-based films functionalized (a) without and (b) with streptavidin captured inside the DNA origami-positioned Rh nanoantennas. Different lines represent different areas on the sample. d-e) Mean UV-SERS spectra with error bars calculated from the data in (a) and (b), respectively. c) Single-protein UV-SERS mapping spectra of a sample containing a low surface density of 24 nm Rh NCs with a single streptavidin protein. Substrate-specific control. f) UV-SERS spectra obtained from repeatable measurements on adjacent aluminum and silica regions of a micropatterned substrate. Inserts: (Top) Optical micrography of the measured Al/Si pattern (scale bar: 4μm). (Down) UV-SERS intensity map at 1061cm$^{-1}$ (characteristic of the DNA phosphate backbone) across a 70 μm$^2$ area, highlighting a silica circle (diameter ≈ 39.8 μm) used for spatial registration.

A detailed vibrational assignment for the range of our observed Raman bands is provided in Table 1. This analysis confirms that the new broad feature bands in the protein-bound spectra originate from characteristic modes of streptavidin. Specifically, peaks assignable to tryptophan residues (e.g., ~1336 and ~1367 cm$^{-1}$)[25,53–55], tryptophan W2 (e.g., ~1460-1580 cm$^{-1}$)[25,53,55], tryptophan Y8b / W1 (e.g., ~1608 and ~1622 cm$^{-1}$)[55] and the amide I band (~1670 cm$^{-1}$)[25,53,55] are clearly identified in Fig. 1(b) and (e), which are absent in the control spectrum dominated by DNA signatures in Fig. 1(a) and (d). This clear spectroscopic distinction provides direct evidence for the specific capture of streptavidin at the ensemble level.

**Table 1** SERS Peak Assignments for DNA-Origami Scaffold and Streptavidin

| Observed Peaks (cm$^{-1}$) | Assignment [a] | | Vibrational Mode | Key References |
|---|---|---|---|---|
| ~650-680 | **DNA** | G | Ring breathing/deformation | 56–58 |
| ~731 | | A | Ring breathing | 26,56–60 |
| ~785 | | T/C | Ring breathing | 56–58 |
| ~1061 | | Phosphate Backbone | PO$_2^-$ symmetric stretch | 26,57,58,60 |
| ~1250 | | C | Ring vibration (C-N stretch) | 26,56,57 |
| ~1334 | | A | Ring vibration (C-N stretch) | 56–60 |
| ~1370 | | T | Symmetric CH$_3$ bending | 56,57 |
| ~1482 | | G/ A/ C | Ring stretching/deformation coupling | 26,56,57 |
| ~1520 | | C | Ring stretching vibration | 56,57 |
| ~1548 | | G | Purine ring stretching vibration | 56,57 |
| ~1580-1590 | | A/G | Ring stretching | 56,57,59,60 |
| ~1660-1680 | | T | C=O/C=N stretch (DNA) | 56,57 |
| ~1470 | **Linker: Biotin** | | Stretching of ring C-H$_2$ | 25,53,54 |
| ~1565 | | | C-H stretching | 25,53,61 |
| ~1336 | **Streptavidin** | | Tryptophan W7 | 25,53–55 |
| ~1367 | | | Tryptophan W7 | 25,53–55 |
| ~1450 | | | CH$_2$/CH$_3$ deformation | 25,53,55 |
| ~1560-1580 | | | Tryptophan W2 | 25,53,55 |
| ~1590 | | | Tryptophan Y8a | 55 |
| ~1608 | | | Tryptophan Y8b / W1 | 55 |
| ~1622 | | | Tryptophan Y8b / W1 | 55 |
| ~1670 | | | Amide I/β-sheeet | 25,53,55 |

[a] A, adenine; T, thymine; G, guanine; C, cytosine

To advance to single-molecule sensitivity, we conducted high-resolution mapping (2400 gr/mm) on a sparsely distributed (low spatial density, i.e. <1 Rh NCs dimer per μm$^2$) sample containing 24 nm Rh-NCs dimers (Fig. 2c). In this case we can report a single point with a detectable spectrum from the protein can be unambiguously assigned. The key protein-specific peaks identified in the ensemble measurements are consistently observed in these single-molecule spectra, confirming that the signal originates from individual streptavidin proteins positioned within the plasmonic hotspot.

These results collectively demonstrate that our architecture not only generates a specific UV-plasmonic response to protein binding but also achieves the spatial and spectral resolution necessary for detecting and spectroscopically identifying individual biomolecular interactions. To further validate the substrate specificity and spatial control of our platform, we also performed integrated UV-SERS mapping on micropatterns comprising alternating stripes of aluminum and silicon (see methods for details). The role of Al as substrate for the Rh-NCs has been discussed in our recent work[39] where we showed how the "mirror" effect of the metallic substrate significantly contributed to increase the electromagnetic field confinement and enhancement (to further support this claim we report additional numerical simulations in the SI – note#11). We tested two assembly densities of DNA-origami-based Rh NCs dimers, corresponding to average inter-dimer spacings of approximately 1 μm ('0.5X') and 2 μm ('0.2X'). As shown in Fig. 2f for the '0.2X' density, characteristic protein UV-SERS signals were exclusively detected on the aluminum stripes, while no measurable signal originated from the adjacent silicon regions. This stark, binary contrast was consistently observed across both densities and was also confirmed for '0.5X' density with and without streptavidin, as illustrated in the SI – Fig. S6. Together, these control experiments provide definitive evidence that the observed UV-SERS enhancement is an intrinsic property of the aluminum-based plasmonic cavity, with no measurable contribution from the silicon substrate. The inset in Fig. 2f provides spatial context: an optical micrograph of the Al/Si pattern and a SERS intensity map at 1061 cm$^{-1}$ (DNA phosphate backbone), confirming the precise localization of the measurement.

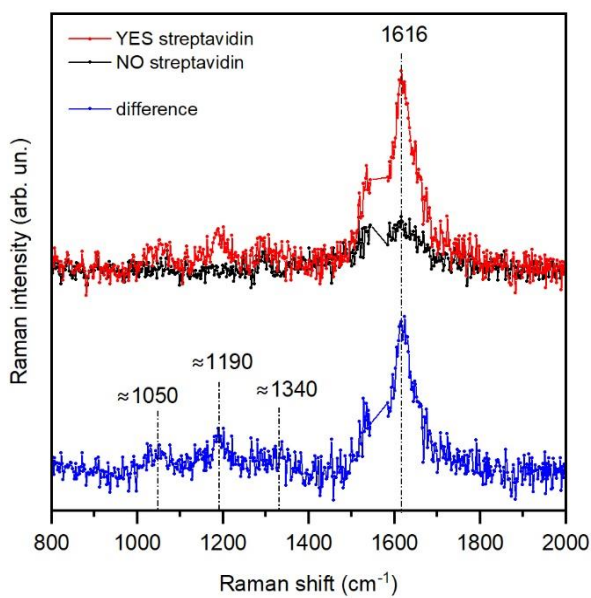

**Fig. 3.** Baseline-subtracted UV-SERS spectra under 266 nm excitation for samples a) with and b) without streptavidin, respectively. Note the distinct peak at around 1616 cm$^{-1}$, characteristic of streptavidin, which appears exclusively in the functionalized sample.

Given the relatively broad spectral features obtained using the excitation wavelength at 325 nm, we further employed a 266 nm laser (setup presented in Fig. S4) to validate the platform's detection performance. Both DNA and some amino-acids present specific absorbance close to this spectral wavelength, therefore it is interesting to compare the SERS performances at this second excitation wavelength. Under this condition, we tested both streptavidin-free and streptavidin-functionalized samples at the '0.5X' density. For these measurements, the sample was rotated (see Fig. S4) to avoid photodegradation. Measurements done without rotation produce significant photodegradation in less than 1 minute (see SI – note#6). The final spectrum represents the average of 12 acquisitions with a 5 minutes integration time each, for a total integration time of 30 minutes. The results are presented in Fig. 3. The black spectrum represents the Control sample ("streptavidin-free") while the red spectrum corresponds to the streptavidin-functionalized sample. The difference between the two spectra is reported in blue, to better enhance the streptavidin spectral features. Results shows that a distinct and prominent characteristic peak at approximately 1616 cm$^{-1}$ is clearly observed in the functionalized sample, which observed the peak around 1622 cm$^{-1}$ in Table 1 assigned to Tryptophan Y8b / W1 (to note Tryptophan is the aromatic amino-acid with the strongest absorbance at 266 nm). At the same time small peaks at 1050, 1190 and 1340 cm$^{-1}$ can be appreciated in the spectrum. These features can be referred to respectively to W1/NO3, Y7A/Y9A/W10 and W7 streptavidin spectral features, recognized in scientific literature trough UV Raman[55]. It provides further confirmation of specific protein detection under optimized UV excitation.

**Validating Polarization-controlled UV-SERS of streptavidin**

The deterministic control over the local electromagnetic field via nanoantenna orientation is a cornerstone of plasmonics, as the excitation of specific plasmon modes depends on the alignment between the incident field and the nanostructure's principal axes[41–44]. To experimentally validate this principle in our DNA-assembled system and quantify its impact on single-molecule detection, we performed comprehensive polarization-resolved UV-SERS measurements using both 325 nm and 266 nm excitation wavelengths (Fig. 4). To note, in order to perform this analysis, samples with low spatial density of Rh-NCs (0.5X density, i.e. <1 nanostructure per μm$^2$) were used. In fact, while the biofabrication ensures the positioning of the dimers on the DNA template, the different structures are then randomly deposited on the Al coated silicon substrate, therefore, only analyzing single structures (and the corresponding single protein) it could be possible to obtain a valid polarization-dependent spectrum. As shown in Fig. 4a, we initially characterized the spectra of a streptavidin-free sample as control, we found the intensity of the peak associated to the DNA structure are reduced by adding a polarizer, with the spectral features only slightly changed with the introduction of the polarizer at different angles, with the intensity of the peak depending on the polarizer orientation. Considering then the streptavidin-functionalized sample, it is possible to observe how the UV-SERS intensity (Fig. 4b) exhibits a strong and systematic modulation as the polarizer is rotated. For reference, the spectrum acquired without any polarizer is provided in Fig. S7 - SI, which shows the highest baseline intensity before polarization-dependent attenuation. Fig. 4c clearly presented the quantitative analysis of the integrated intensities for three diagnostic peaks (1607, 1621, and 1660 cm$^{-1}$) extracted from Fig. 4b. It revealed a clear characteristic of a dipolar plasmonic response. Critically, the maximum intensity for all peaks occurred when the incident electric field was polarized at 22.5°—parallel to the long axis of the NCs dimer—with a minimum at the orthogonal direction (0°) in Fig. 4c. This dependence directly correlates the external optical field with the excitation efficiency of the dimer's fundamental mode, as predicted by simulations in literature[41].

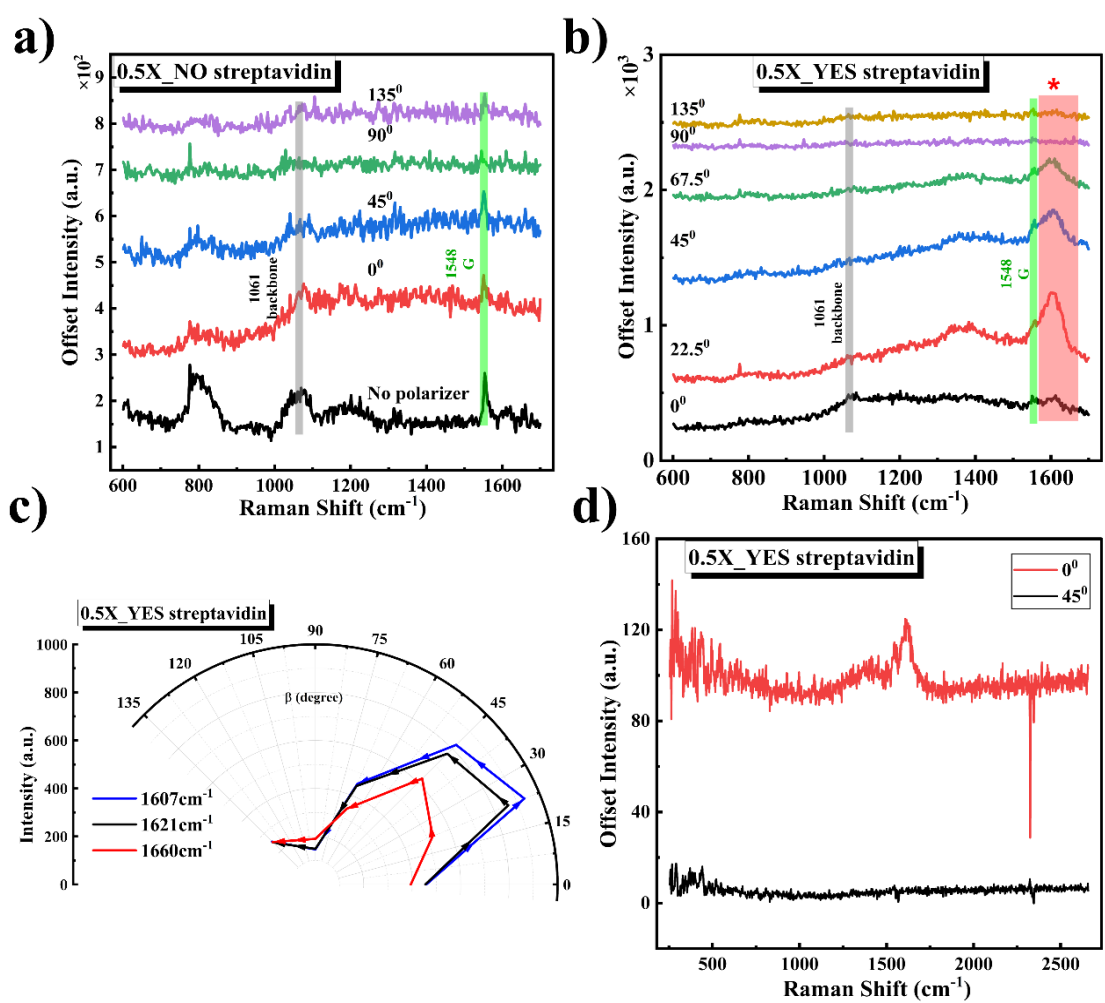

**Fig. 4** Polarized UV-SERS spectra of single streptavidin on DNA Origami under 325nm and 266nm excitation. a-d) 325nm excitation: a) UV-SERS spectra of the streptavidin-free control

sample without a polarizer (black) and with the polarizer rotated from 0° to 135° (color gradient from red to purple). b) Corresponding SERS spectra of the streptavidin-functionalized sample at different polarizer angles. c) Integrated peak intensities at 1607, 1621, and 1660 cm$^{-1}$ (blue, black, and red symbols, respectively) extracted from the spectra in b). d) 266nm excitation: UV-SERS spectra of the streptavidin-functionalized sample under 0° and 45°polarized excitation, respectively.

Polarization measurements performed also using the excitation wavelength at 266 nm. These second set of measurements confirmed the orientation-dependent enhancement (Fig. 4d). The spectrum acquired without a polarizer in Fig. 3 showed an intermediate intensity, representing an incoherent average, while the significant modulation obtained with the adjunct of the polarized in Fig. 4d directly shows the anisotropy of the plasmonic hotspot engineered by the well-defined dimer geometry. Collectively, these data provide the first experimental validation of polarization-selective UV-SERS in a deterministically assembled DNA-origami nanoantenna. We demonstrate that the SERS intensity of a single protein is actively modulated by incident polarization, with maximum enhancement achieved at alignment with the dimer's principal axis (0°) under 266nm excitation. This establishes our platform as offering not only single-molecule sensitivity but also directional control over the enhancement, a critical advancement for tailored nanophotonic sensing.

## Discussion

Our work establishes a programmable platform that merges DNA nanotechnology (biofabrication) with UV plasmonics to achieve single-protein detection via label free Raman spectroscopy. We moved beyond conventional gold-based systems by integrating Rh NCs, whose UV resonances are crucial for exciting biomolecular electronic transitions, into a geometrically precise DNA origami scaffold. This assembly strategy, a decisive step beyond earlier stochastic methods, reliably creates "hot spots" while enabling the deterministic placement of a target molecule.

The observation of characteristic Raman bands from a single streptavidin protein, such as tryptophan and amide vibrations, confirms the platform's specificity and single-molecule sensitivity in the UV range. More importantly, we provide the first experimental demonstration of polarization-dependent control in such a DNA-assembled UV nanoantenna. The modulation of signal intensity with incident polarization—and the clear maximum at a specific angle—directly validates the coupling between the external optical field and the engineered plasmonic mode of the dimer, transforming the structure from a passive amplifier into an active optical element.

These capabilities are fundamentally enabled by a synergistic plasmonic cavity. Our micropatterned control experiments offer visual proof that the aluminum film substrate is indispensable, actively contributing to the enhancement rather than serving as a passive support. This synergy among material (Rh), geometry (dimer), and substrate (Al) forms the core of our platform's performance.

Looking forward, the inherent programmability of DNA origami provides a clear path for advancement. The scaffold can be redesigned to tune the nanoantenna's spectral response or to present multiple probes for multiplexed detection. The proven polarization control further suggests promising routes toward developing chiral plasmonic sensors. While challenges remain in areas like large-scale uniformity and absolute enhancement factor quantification, this work decisively extends the reach of DNA-assembled plasmonics into the UV regime. It lays a versatile foundation for applications demanding label-free, single-molecule spectral fingerprinting with built-in optical tunability, from fundamental bio-spectroscopy to next-generation sensing.

## Methods

### Rhodium nanoparticles synthesis

Rhodium nanocubes were synthesized using a seed-mediated polyol method following Zhang et al. (Size-Tunable Rhodium Nanostructures for Wavelength-Tunable Ultraviolet Plasmonics, Nanoscale Horizons, 2015). Potassium bromide and polyvinylpyrrolidone were used as shape-directing and stabilizing agents, with ethylene glycol acting as solvent and reducing agent. Seed particles were first prepared at 160 °C, followed by a controlled growth step via injection of a rhodium precursor into the seed solution. The reaction was maintained at 160 °C and then cooled to room temperature, yielding rhodium nanocubes with average edge lengths of approximately 24 and 40 nm. After synthesis, the nanocubes were washed and redispersed in deionized water.

### Assembly of DNA Origami "Boat"

DNA origami structures were assembled from a 7249-nt M13mp18 scaffold using 243 staple strands in 1× TAE buffer supplemented with 12 mM $MgCl_2$ (pH 8), following the protocol described in DNA-Origami-Assembled Rhodium Nanoantennas for Deep-UV Label-Free Single-Protein Detection (Corduri et al.). Staples were added in 10-fold excess over the scaffold (100-fold for functional staples). The mixture was thermally annealed from 70 °C to 25 °C. Purification was performed either by agarose gel electrophoresis or by Amicon ultrafiltration. Streptavidin was subsequently added in 20-fold excess, incubated at room temperature, and the final constructs were purified by agarose gel electrophoresis and stored at 4 °C.

<u>Rhodium dimers assembly</u>

Rhodium nanocube dimers were assembled by incubating T18-functionalized rhodium nanocubes with DNA origami at a 20:1 ratio in 1× TAE buffer containing 12 mM $MgCl_2$ and 600 mM NaCl, following the procedure described in Corduri et al., DNA-Origami-Assembled Rhodium Nanoantennas for Deep-UV Label-Free Single-Protein Detection. The mixture was incubated overnight at room temperature, and the resulting dimers were purified by 1% agarose gel electrophoresis. The corresponding band was extracted and used without further purification.

<u>Deposition onto Si/Al substates</u>

To prepare the substrates for the Rh NCs deposition, we used 1x1 cm double polished Si wafer (500 μm thick) pieces. After proper cleaning (successive washing in acetone, isopropanol and water), the Si surface was coated with 20 nm of Al evaporated by means of e-beam in high-vacuum. To facilitate the comparison of the SERS performance on Al and on Si, patterned substrates were prepared via simple optical lithography creating small circles of Al over the Si surface (see Fig. 1f). Rh NCs dimers (160 pM) were deposited onto aluminum coated (or on Si for control) substrates by incubation for 30 min at room temperature. After deposition, the substrates were gently rinsed with milli-Q water to remove unbound structures and used for subsequent measurements.

## TEM measurements

For TEM imaging, 5 uL of the solution were dropped onto EM-Tec Formvar Carbon support film on copper 300 square mesh. After 2 min the solution was removed with a paper filter, and the sample was stained using 2% Uranyl Acetate for 40 sec followed by a quick water rinsing. The measurements were then carried out using a Tecnai Spirit with an accelerating voltage of 120kV.

## Optical microscopy and spectroscopy

-SERS measurements at 325 nm were performed using a HORIBA LabRAM HR Evolution Raman spectrometer (Horiba Jobin Yvon, Kyoto, Japan) with 40X UV objective integrated Raman system couples a confocal microscope to an 800mm spectrograph with two switchable gratings. The UV-SERS spectra were conducted using 325nm laser with 500-hole value. Fig. 2a and 2b were obtained by using a 1800gr/mm grating to scan 70μm$^2$ area, 14μm step size and total 36 points, every point collected with 270s exposure time, and one accumulation at 10% laser power attenuation. Fig. 2f were using the same parameter with a 1800gr/mm grating and 270s exposure time, and one accumulation at 10% laser power attenuation as Fig. 4a and 4b. But Fig. 2c were acquired by 2400gr/mm grating with 100μm$^2$ area, 5μm step size and total 441 points, every point collected with 120s exposure time, and one accumulation at 5% laser power attenuation.

-UV Raman measurements were performed using a 266 nm excitation wavelength. Diffused light was collected in backscattering configuration by means of a 2-inch plano-convex lens with a focal length of 10 cm. The beam power was set to 0.2 mW to minimize sample photodegradation as much as possible. A Czerny-Turner spectrometer (Andor 750), with f/9.4 and a focal length of 750 mm, coupled with a back-illuminated, Peltier-cooled CCD (Andor Idus), was used to obtain the final Raman spectra. The spectral resolution was set to 20 $cm^{-1}$ to maximize the count rate. During some measurements, the sample was continually rotated, keeping the focus position on the sample surface, to avoid photodegradation. The final spectra were obtained as the average of 12 spectra of 5 minutes each. For the polarization measurements, a half-lambda waveplate was used to set the polarization of the incident light with respect to the sample orientation. Two polarizations were used, namely VV, with the electric field parallel to the sample edge, and V45, with the electric field rotated by 45° with respect to the sample edge. In this case sample was not rotated and final were obtained as the average between 5 spectra of 1 minute each, collected in different points.

**Author Contributions**

YZ and FDA performed the SERS measurements, NC and KK prepared the origami samples, HJ, AD, SW supported with the samples fabrication, YY, JL and JW provided the Rh NCs, SJ, ZZ, GA and DG supervised the work.

**Notes**

The authors declare no competing financial interest.

# Acknowledgements

The authors thank the European Union under the HORIZON-Pathfinder-Open: 3D-BRICKS, grant Agreement 101099125; DYNAMO, grant Agreement 101072818.

# Supporting Information

## Sel-assembled Rhodium Nanoantennas for Single-Protein UV SERS

*Yanqiu Zou[1], Nicco Corduri[2], Francesco D'Amico[3], Karol Kolataj[2*], Huaizhou Jin[4], Zhenrong Zheng[1*], Yifan Yu[5], Jie Liu[5], Shukun Weng[6], Ali Douaki[6,7], Jerome Wenger[8], Shangzhong Jin[9*], Guillermo Acuna[2*], Denis Garoli[6,7,9*]*

[1] State Key Laboratory of Modern Optical Instrumentation, College of Optical Science and Engineering, Zhejiang University, Hangzhou 310027, China

[2] Department of Physics, University of Fribourg, Chemin du Musée 3, Fribourg CH-1700, Switzerland

[3] Elettra Sincrotrone Trieste S.C.p.A., S.S. 14 km 163,5 in Area Science Park, 34149,

Basovizza, TS, Italy

[4] Key Laboratory of Quantum Precision Measurement, College of Physics, Zhejiang University of Technology, Hangzhou, China

[5] Department of Chemistry, Duke University, Durham, NC 27708, USA

[6] Istituto Italiano di Tecnologia, via Morego 30, I-16163, Genova, Italy

[7] Dipartimento di Scienze e Metodi dell'Ingegneria, Università degli Studi di Modena e Reggio Emilia, Via Amendola 2, 42122, Reggio Emilia (Italy)

[8] Aix Marseille Univ, CNRS, Centrale Marseille, Institut Fresnel, AMUTech, 13013 Marseille, France

[9] College of Optical and Electronic Technology, China Jiliang University, Hangzhou 310018, China

**Note #1 – TEM characterization of Rh NCs**

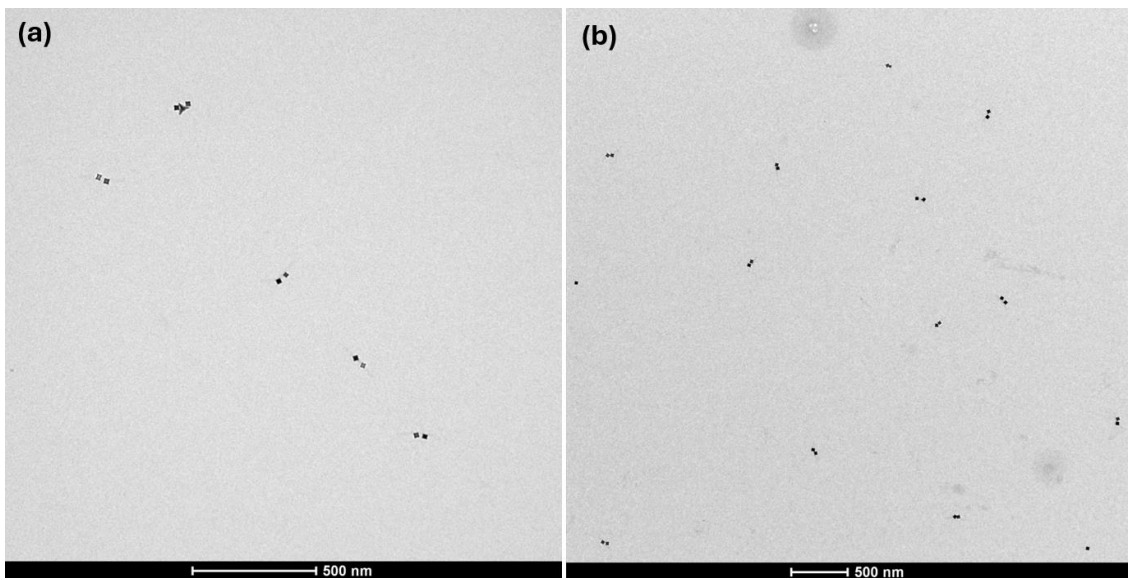

**Fig. S1.** TEM micrographs of (a) 19 nm Rh NCs dimers, (b) 24 nm Rh NCs dimers.

**Note #2 – Absorbance spectra of the Rh NCs**

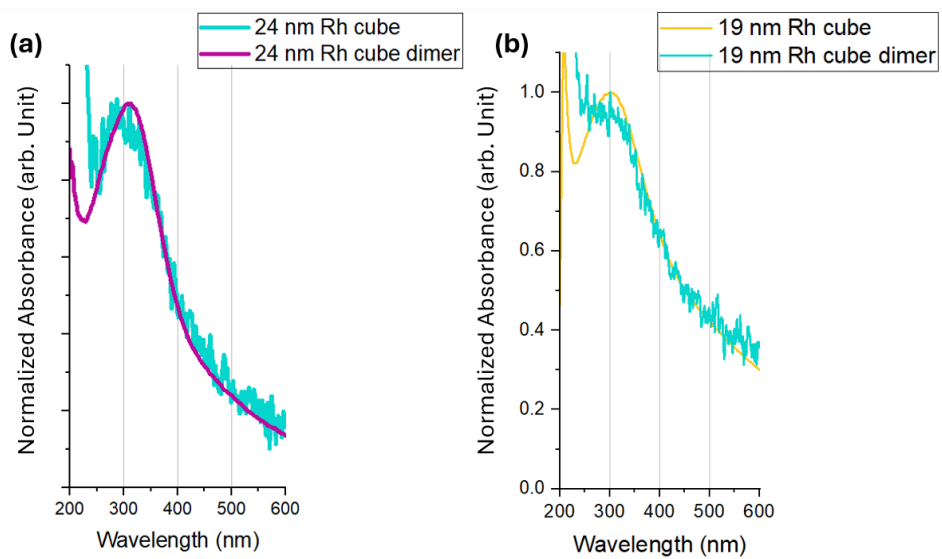

**Fig. S2.** Absorbance spectra of the Rh NCs and Rh dimers. (a) Rh NCs with a size of 24 nm; (b) Rh NCs with a size of 19 nm.

**Note #3 – Raman setup**

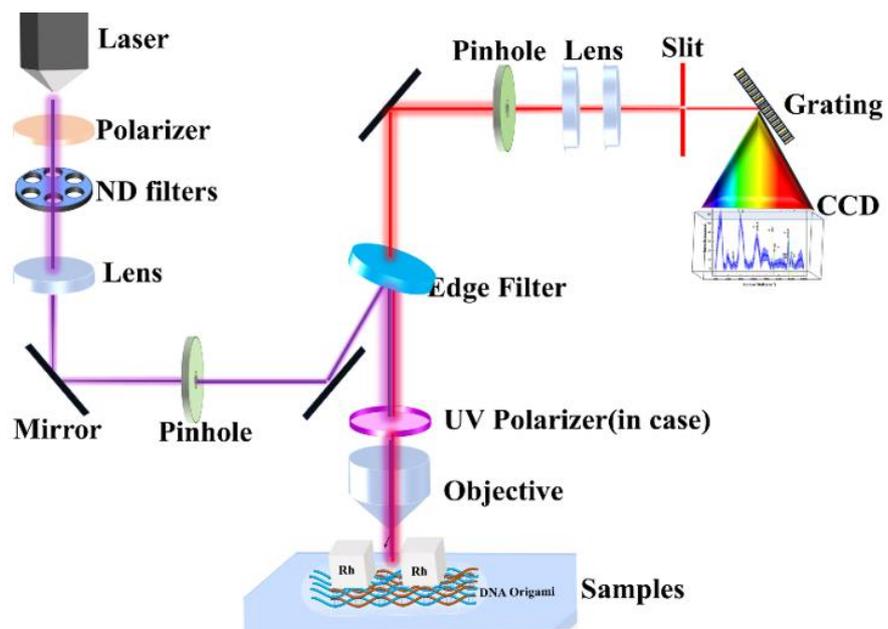

**Fig. S3.** Raman spectrometer setup for the measurements at 325 nm.

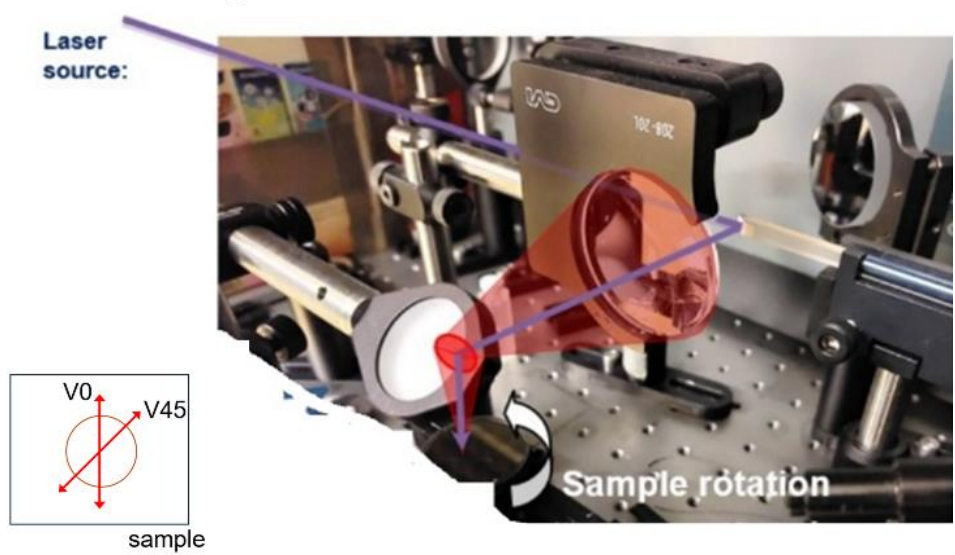

**Fig. S4.** Details on the Raman spectrometer for the measurements at 266 nm – sample+polarizer holders.

**Note #4 – Origami + Rh NCs configurations**

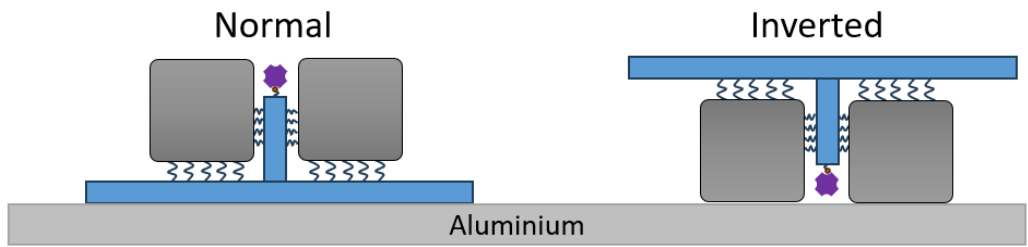

**Fig. S5.** Potential configurations of the origami+Rh NCs after the drop cast on the Al substrate.

**Note #5 - Substrate-Specific Control on Micropatterned Substrates**

To rigorously decouple the plasmonic enhancement originating from the aluminum film from any non-specific background, we performed integrated UV-SERS mapping on engineered micropatterns. These patterns consisted of alternating, well-defined stripes of aluminum (plasmonically active) and silica (plasmonically inactive), as shown in Fig. S6a (inset).

We tested the system at a '0.5X' DNA-origami assembly density (average inter-dimer spacing ~1 μm). The samples were prepared both with and without the target protein, streptavidin. The full set of correlative characterization data—including large-area SERS maps, spatially resolved intensity profiles of the characteristic DNA band at 1061 cm$^{-1}$, and high-resolution AFM topographs—is compiled in Fig. S6.

The results demonstrate a definitive, binary response: characteristic SERS signals (from either the DNA scaffold or the captured protein) were exclusively and reproducibly detected on the aluminum stripes. No measurable signal above the noise floor was obtained from the adjacent silica regions. This clear spatial contrast provides direct and visual evidence that the observed UV-SERS enhancement is an intrinsic property of the aluminum-based plasmonic nanocavity and is not supported by the silica substrate.

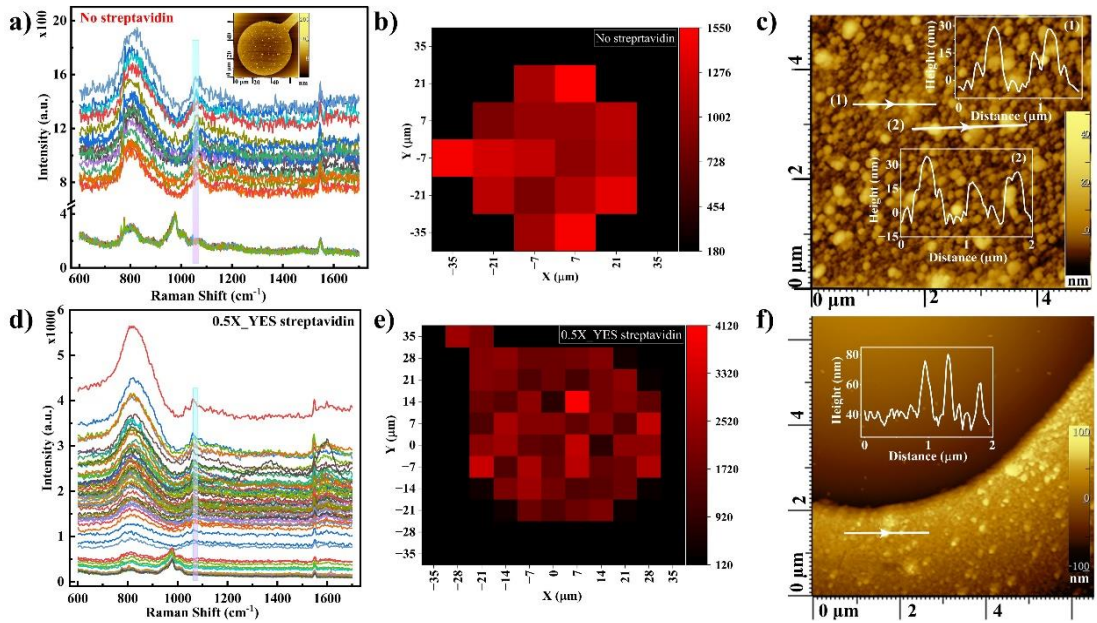

**Fig. S6.** Substrate-specific UV-SERS response on Al/Si micropatterns for the '0.5X' DNA-origami assembly density. a-c) Sample without streptavidin: a) UV-SERS mapping spectra of the DNA-origami scaffold. Inset: AFM image of the measured area (70 μm × 70 μm). b) Corresponding SERS intensity map derived from the 1061 cm$^{-1}$ band (DNA phosphate backbone). c) High-resolution AFM image (5 μm × 5 μm) of the sample surface. d-f) Sample functionalized with streptavidin: d) UV-SERS mapping spectra. e) SERS intensity map at 1061 cm$^{-1}$. f) Corresponding high-resolution AFM image (5 μm × 5 μm).

**Note #6 - UV-SERS spectra without polarizer**

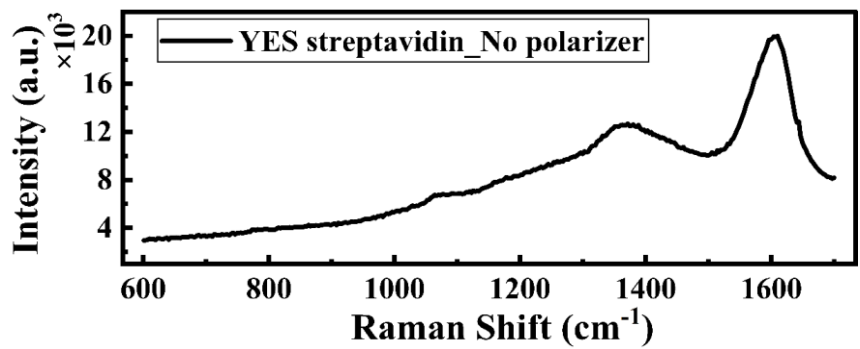

**Fig. S7.** UV-SERS spectra of the streptavidin-functionalized sample under 325 nm excitation without a polarizer.

Fig. S2 presents the UV-SERS spectrum of the streptavidin-functionalized sample acquired without a polarizer. This spectrum serves as a reference and demonstrates that the signal intensity is significantly reduced when a polarizer is introduced and rotated, as shown in the polarization-dependent measurements in Fig. 4a of the main text.

**Note #7 - Photodegradation tests at 266 nm**

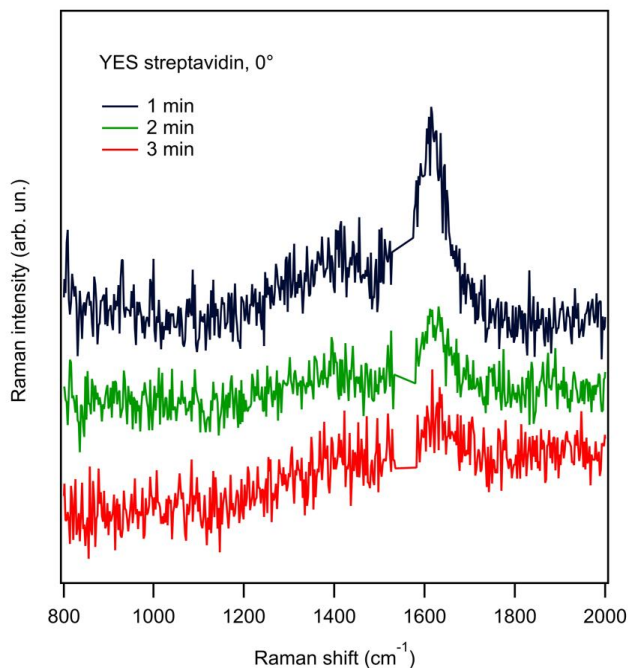

**Fig. S8.** Photodegradation tests at 266 nm excitation wavelength.

**Note #8 - Control samples on Silicon Substrates**

To assess the specificity and structural role of the DNA-origami scaffold in our UV-SERS system, control measurements were performed. As shown in Fig. S3a, the silica-only sample serves as a reference, the DNA-origami-assembled Rh-nanoparticle dimer sample (Fig. S3b) not only shows enhanced SERS signals due to the well-defined plasmonic architecture, but also presents a distinct DNA-related vibrational peak at approximately 1555 cm$^{-1}$, which is absent in the silica reference. This peak confirms the presence and structural integrity of the DNA scaffold in the assembled nanoantenna.

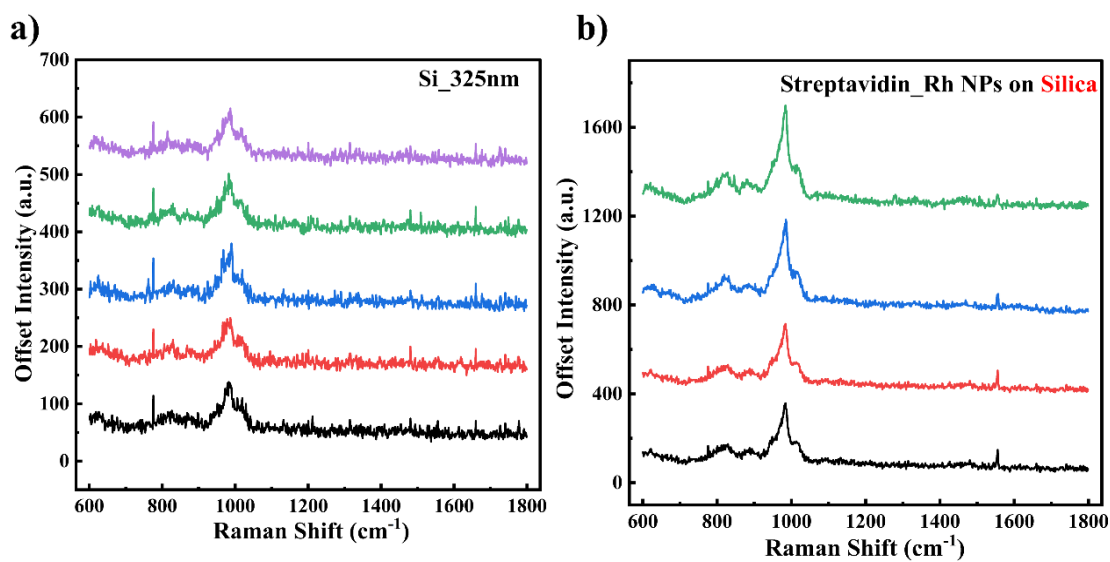

**Fig. S9** UV-SERS spectra of a) silica and b) DNA Origami with Rh NCs functionalized with streptavidin on silica substrates. Spectra were collected from randomly selected points under 325 nm excitation (40× NUV objective). For signal optimization, laser filters were set to 10 % for silica and 50 % for DNA-origami samples.

To systematically evaluate the contribution of the Rhodium nanoparticles and the streptavidin functionalization to the observed SERS enhancement, control SERS mapping experiments were performed under 325 nm excitation (Fig. S4). As shown in Fig. S4a–c, three control samples were examined: DNA-origami with streptavidin and Rh nanoparticles (a), the same structure without Rh nanoparticles (b), and bare DNA-origami on silica (c). The corresponding averaged SERS spectra (Fig. S4d) reveal distinct spectral profiles. The sample containing Rh nanoparticles (red line) exhibits pronounced enhancement across characteristic bands, while the sample without nanoparticles (blue line) and the bare DNA-origami (green line) show significantly weaker signals, comparable to the silica substrate background (black line). These results confirm that the SERS enhancement is primarily mediated by the plasmonic activity of the Rhodium nanoparticles, and that the streptavidin functionalization, in the absence of nanoparticles, does not yield measurable SERS activity under these experimental conditions.

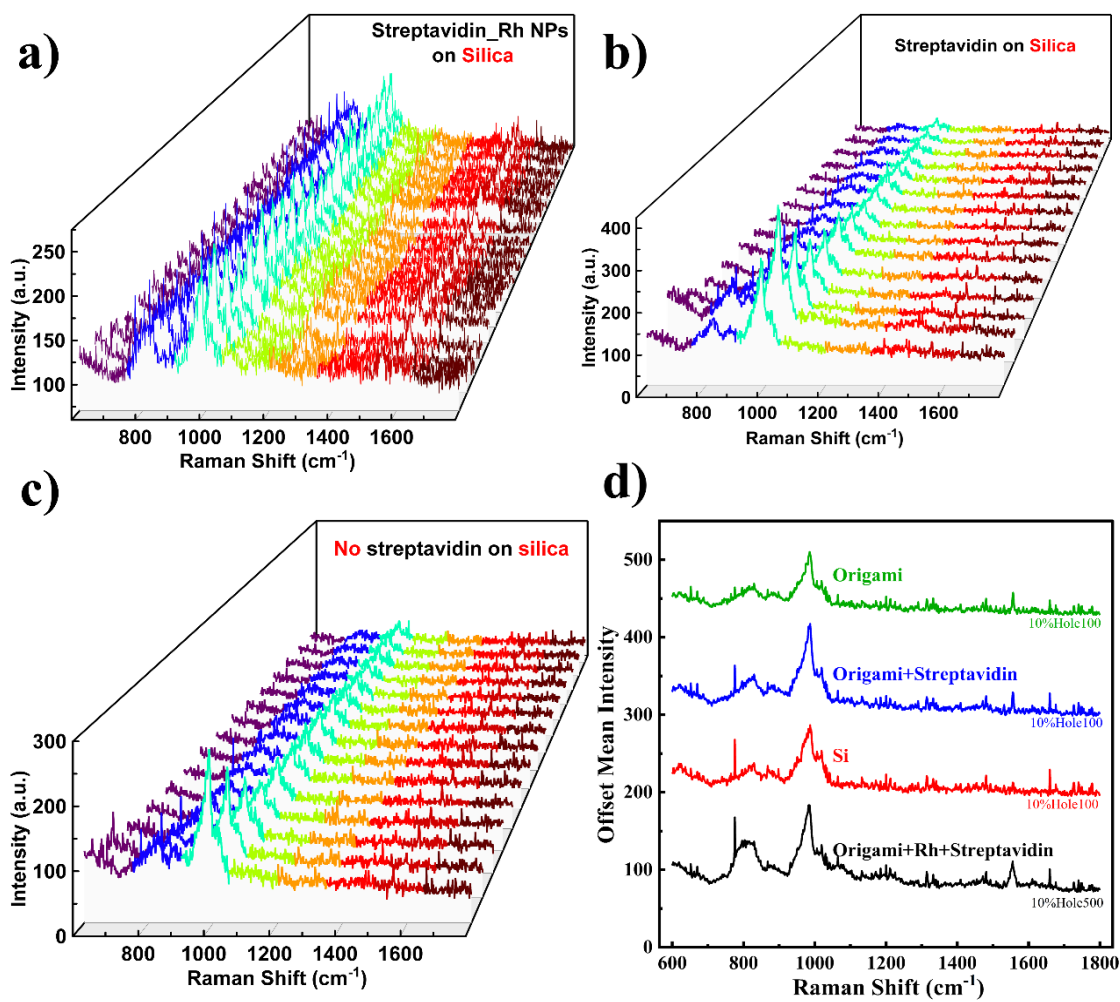

**Fig. S10** UV-SERS mapping of control samples. SERS mapping performed under 325 nm excitation with a 10% laser power filter and 1800 gr/mm grating for the following samples: a) DNA-origami functionalized with streptavidin and assembled with Rhodium nanoparticles on silica; b) the same structure without Rhodium nanoparticles; c) bare DNA-origami on silica substrate. d) Averaged UV-SERS spectra extracted from the regions marked in panels a–c), displayed as red, blue, and green lines respectively, along with the spectrum of a plain silica substrate (black line) for reference. Mapping was conducted with a 500 μm pinhole for a) and a 100 μm pinhole for b) and c).

**Note #9 – Numerical Simulations**

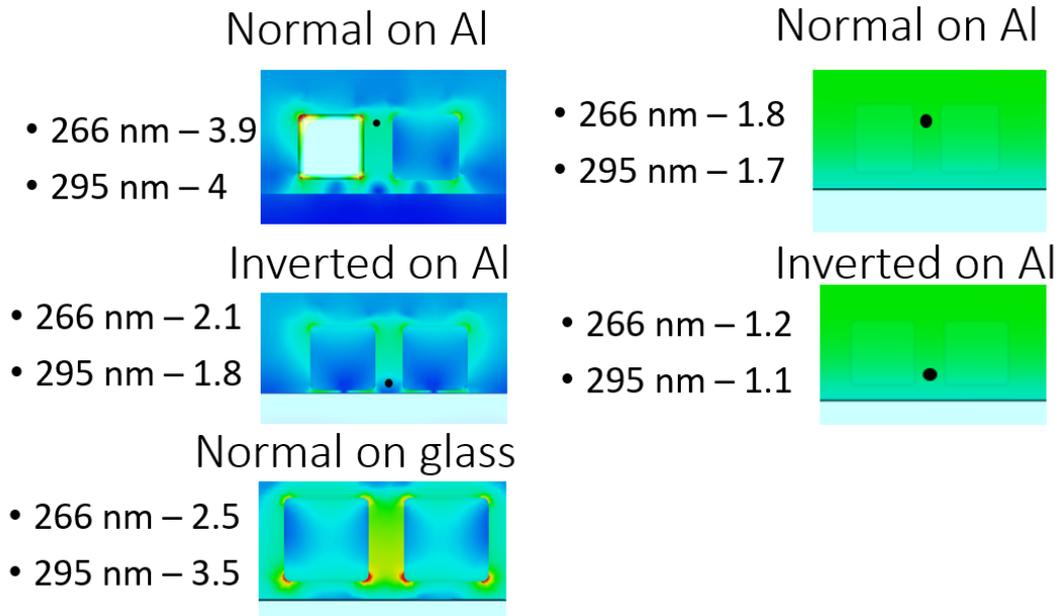

Fig. S11 Numerical Simulations